\documentclass[11pt]{article}
\usepackage[utf8]{inputenc}
\usepackage[T1]{fontenc}
\usepackage[english]{babel}
\usepackage[margin=1in]{geometry}
\usepackage{times}
\usepackage{setspace}
\usepackage{authblk}
\usepackage{amsmath}
\usepackage{amssymb}
\usepackage{amsfonts}
\usepackage{amsthm}
\usepackage{mathtools}
\usepackage{graphicx}
\usepackage{xcolor}
\usepackage{adjustbox}
\usepackage[figuresleft]{rotating}
\usepackage{pdflscape}
\usepackage{tikz}
\usepackage{booktabs}
\usepackage{array}
\usepackage{tabularx}
\usepackage{longtable}
\usepackage{multirow}
\usepackage{makecell}
\usepackage{tablefootnote}
\usepackage{multicol}
\usepackage{caption}
\usepackage{subcaption}
\usepackage{float}
\usepackage{placeins}
\usepackage{algorithm}
\usepackage{algpseudocode}
\usepackage{enumitem}
\usepackage{footnote}
\usepackage{newclude}
\usepackage{xspace}
\usepackage{pifont}
\usepackage{microtype}
\usepackage[none]{hyphenat}
\usepackage[numbers,sort&compress]{natbib}
\usepackage{url}
\usepackage{hyperref}
\hypersetup{
    pdfstartview=FitH,
    colorlinks=true,
    urlcolor=blue,
    citecolor=blue,
    linkcolor=blue
}
\usepackage{cleveref}

\definecolor{lime}{HTML}{A6CE39}
\DeclareRobustCommand{\orcidicon}{
	\begin{tikzpicture}
	\draw[lime, fill=lime] (0,0) 
	circle [radius=0.16] 
	node[white] {{\fontfamily{qag}\selectfont \tiny ID}};
	\draw[white, fill=white] (-0.0625,0.095) 
	circle [radius=0.007];
	\end{tikzpicture}
	\hspace{-2mm}
}

\foreach \x in {A, ..., Z}{
	\expandafter\xdef\csname orcid\x\endcsname{\noexpand\href{https://orcid.org/\csname orcidauthor\x\endcsname}{\noexpand\orcidicon}}
}

\title{Differentially Private EEG Feature Anonymization: A Privacy--Utility Case Study in Clinical Neurophysiology}

\author{Noman Sadiq\orcidA{} \hspace{4cm} Mohsen Toorani\orcidB{} \\
\vspace{10pt}
Department of Science and Industry Systems \\ University of South-Eastern Norway \\ Kongsberg, Norway}
\date{}

\begin{document}
\maketitle
\let\thefootnote\relax
\footnotetext{\scriptsize{Copyright \copyright~Authors 2026. All Rights Reserved. This preprint is distributed through arXiv under the arXiv.org perpetual, non-exclusive license. No additional reuse rights are granted except as provided by that license.}}

\begin{abstract}
Clinical electroencephalography (EEG) data are valuable for healthcare research and for developing artificial intelligence (AI)-based clinical decision-support systems, but EEG recordings and derived features may contain sensitive patient-specific information. This creates privacy risks when data are reused, analyzed, or shared across clinical and research environments. Conventional anonymization methods are often insufficient for high-dimensional biomedical signals, since removing direct identifiers does not necessarily prevent re-identification, linkage, or inference risks. At the same time, strong privacy protection may distort clinically relevant signal characteristics and reduce data utility. This paper studies subject-level differential privacy for protecting clinical EEG-derived feature representations using Gaussian and Laplace perturbations. The proposed framework considers three deployment scenarios: client-side anonymization, centralized server-side anonymization, and decentralized local training. Following EEG preprocessing and feature extraction, Gaussian and Laplace perturbations are applied to the resulting patient-level EEG feature representations. The Laplace experiments evaluate the implemented noise scales, while the scales required for formal full-vector calibration are derived separately. The effects of both perturbations are assessed using statistical utility measures and a downstream machine-learning-based utility check. The results show that differentially private perturbation can be integrated into EEG processing workflows, but the selected mechanism, privacy parameters, and sensitivity calibration strongly influence data utility. The study highlights the practical privacy--utility trade-off in DP-based EEG feature anonymization and the challenges of preserving downstream utility in small and imbalanced clinical EEG datasets.
\end{abstract}

\section{Introduction}
Healthcare increasingly depends on digital technologies for collecting, storing, and analyzing patient data to improve efficiency, accuracy, and quality of care. At the same time, the growing collection and sharing of healthcare data creates serious privacy challenges. Patient records are highly sensitive, and healthcare institutions are under increasing pressure not only to use data for direct treatment but also for secondary purposes such as clinical research, decision support, and training of artificial intelligence (AI) models. This creates a need for privacy-preserving data sharing mechanisms that can support innovation while still meeting legal and ethical requirements. If privacy protection is weak, patient trust may be reduced, and organizations may become more reluctant to share valuable data. On the other hand, if privacy protection is too strict, important medical knowledge may be lost. Therefore, the balance between privacy protection and data utility has become a central problem in healthcare informatics \citep{introduction-1}.

This challenge is especially important for clinical neurophysiological data such as EEG (Electroencephalography) and electromyography (EMG). Unlike many traditional clinical datasets, EEG and EMG data are not only tabular records but also high-dimensional time-series signals collected over longer periods for each patient. These signals contain complex temporal and frequency patterns that are important for diagnosis and machine learning tasks, but they may also reveal sensitive neurological and medical information. As a result, privacy protection for such data is more difficult than simply removing names or patient identifiers \citep{healthcare-intro}. Traditional de-identification models such as $k$-anonymity seek to make each record indistinguishable from at least $k-1$ other records with respect to selected quasi-identifiers \cite{sweeney2002kanonymity}.

Traditional anonymization methods, including suppression, generalization, and $k$-anonymity-based models, provide useful baselines for privacy-preserving data publishing, but they are often insufficient for high-dimensional biomedical signals. Such methods may remain vulnerable to re-identification, linkage, or attribute-inference risks and may also reduce the utility of time-series data when strong generalization or suppression is applied. Differential Privacy (DP) \cite{Dwork2006, DworkRoth} offers a stronger formal framework by adding calibrated randomness to data or query outputs, thereby limiting the influence of an individual patient on the released result. In the present work, DP is considered at the subject level and is applied to EEG-derived feature representations rather than directly to raw EEG recordings. This design is motivated by the need to support secondary analysis and machine-learning use cases while reducing patient-level disclosure risk.

This paper focuses on privacy-preserving processing and sharing of clinical EEG data using Gaussian and Laplace perturbations within a subject-level DP framework. The study evaluates how these perturbations affect data utility and downstream machine learning performance in a clinical EEG case study. The main contributions of this paper are as follows:
\begin{itemize}
    \item A clinical EEG data anonymization case study is presented using subject-level DP applied to EEG-derived feature representations.
    \item Three deployment scenarios are considered for multi-hospital EEG processing: client-side anonymization, centralized server-side anonymization, and decentralized local training.
    \item Gaussian and Laplace perturbation mechanisms are implemented and compared after EEG preprocessing and feature extraction, together with a separate derivation of the full-vector $L_1$ sensitivity and corresponding Laplace scales required for formal calibration.
    \item The impact of the selected Gaussian and Laplace configurations is assessed using statistical utility measures and a downstream machine-learning-based utility check.
    \item Practical challenges are discussed for applying formal privacy mechanisms to small and imbalanced clinical EEG datasets.
\end{itemize}

The scope of the study is limited to selected DP configurations and EEG-derived feature representations. The classification experiment is used as a downstream utility check rather than as the main objective of the paper. The results should therefore be interpreted in light of the limited dataset size, class imbalance, selected sensitivity/noise calibrations, and the absence of extensive empirical attack-based privacy testing. Larger datasets, broader privacy-budget exploration, rigorous privacy accounting, dedicated re-identification, linkage, membership-inference, and reconstruction analyses, and broader model benchmarking are left for future work.

\section{Related Work}

Privacy-preserving EEG analysis is an emerging research area. EEG data can reveal subject identity, neurological characteristics, cognitive states, and other sensitive information, creating risks in brain-computer interfaces, clinical neurophysiology, and collaborative model training. These risks are especially important in healthcare settings where EEG data may need to be reused for research, shared between institutions, or used for machine learning model development. Existing work on privacy-preserving EEG analysis can be grouped into three partly overlapping directions: formal privacy mechanisms such as differential privacy, federated or distributed learning approaches, and representation-learning methods that aim to suppress identity-related information while preserving task utility.

Several studies have applied differential privacy to healthcare and medical data analysis. Subramanian \citep{Subramanian22} studied the Laplace mechanism for healthcare data and discussed how noise affects analytical accuracy. Sun et al. \citep{8868084} proposed a DP-based medical data publishing and training framework using normalization, weighted Laplace noise, gradient clipping, and Gaussian noise to improve privacy while preserving utility. Letafati and Otoum \citep{10294469} considered distributed healthcare settings in which local models are clipped and perturbed with Gaussian noise before aggregation. These studies show that DP can provide a formal privacy framework for medical data analysis, but they also demonstrate the central challenge of balancing privacy protection against data utility. 
Recent work on medical deep learning has further examined the methodological choices, privacy--utility trade-offs, and deployment implications of differential privacy in healthcare applications \cite{mohammadi2026medicaldp}. It is important to distinguish DP applied to a released feature table from DP applied during model optimization. DP-SGD, for example, clips and perturbs per-example gradients and accounts for the privacy loss accumulated over repeated optimization steps, whereas the present study examines a one-time patient-level feature release \cite{Abadi2016,NIST800226}. These settings protect different released objects and require different sensitivity and privacy-accounting arguments. More broadly, decentralized health-data sharing involves several complementary security and privacy requirements beyond perturbation of the released data, including access control, accountability, consent management, and protection during cross-institutional collaboration \cite{NaeemT2026a}.

Machine learning methods are central to EEG analysis because EEG signals are high-dimensional, non-stationary, and subject-dependent. In privacy-preserving EEG research, machine learning is used both for classification and for evaluating whether transformed EEG data retain useful task information. Luo et al. \citep{LuoJQFZ25} applied federated learning with differential privacy to EEG-based epilepsy recognition. Their framework keeps raw EEG data on local clients, extracts EEG features locally, trains local models, and shares only model parameters with the server. Privacy is supported through gradient clipping and Gaussian noise added to model parameters before aggregation. This work is closely related to the present paper because it combines EEG-based epilepsy recognition, federated learning, and DP. However, it focuses on model-parameter perturbation in a federated learning setting, whereas the present paper evaluates Gaussian and Laplace perturbation of patient-level EEG-derived feature representations. More generally, federated learning enables participating institutions to train a shared model while retaining their raw clinical data locally \cite{rieke2020federated}. Nevertheless, gradients and model updates can leak information about the underlying training data through attacks such as gradient inversion \cite{hatamizadeh2023gradient}. Secure aggregation addresses part of this risk by allowing the aggregation server to recover an aggregate of the participating clients' updates without observing each individual update in plaintext \cite{bonawitz2017secureaggregation}. In a clinical EEG setting, Rajabi and Toorani \cite{RajabiT26} implemented and evaluated masking-based secure aggregation with dropout recovery, malicious-setting safeguards, and auxiliary-notary-based verifiability for cross-silo federated learning.

Another line of work focuses on making EEG data less identifiable while preserving task performance. Meng et al. \citep{10236508} studied sample-wise and user-wise perturbations that add near-imperceptible noise to EEG data in order to reduce identity leakage while maintaining BCI task utility. Chen et al. \citep{Chen_2025} also studied user-wise perturbations for identity protection in EEG-based BCI systems, including random, synthetic, error-minimizing, and error-maximizing perturbations. These approaches are useful for reducing identity leakage against specific models or attack settings, but they do not provide the same type of formal privacy guarantee as differential privacy.

Representation-learning approaches have also been proposed for privacy-aware EEG analysis. Singh et al. \citep{10320167} used a multi-objective autoencoder to suppress subject identity while maintaining task recognition. Wang et al. \citep{ijcai2025p469} proposed an identity-removal network that separates task-related and identity-related EEG features and removes identity-related components before decoding. Bethge et al. \citep{bethge2022domaininvariantrepresentationlearningeeg} introduced domain-aligned private encoders for EEG-based emotion recognition, where private encoders and a domain-alignment loss are used to support cross-dataset learning while reducing domain-specific information. These methods are important because they address the fact that EEG contains identity-related structure, but most of them focus primarily on empirical identity suppression rather than formal DP accounting.

Privacy-preserving and data-sharing issues in EEG and BCI have also been studied from a broader methodological perspective. Xia et al. \citep{9760113} proposed augmentation-based source-free adaptation for motor-imagery EEG classification, where the source EEG data are not shared during adaptation to a target user. Although this method is not a DP mechanism, it is relevant because it addresses the practical problem of adapting EEG models without transferring raw EEG data. Xia et al. \citep{XiaDSXFLZSXWW23} reviewed privacy protection in brain-computer interfaces, with a strong focus on EEG data. They summarized data-level and model-level privacy risks and discussed defense strategies such as anonymization, cryptography, differential privacy, federated learning, and transfer learning. Their review highlights the need for clearer privacy--utility evaluation protocols and more standardized ways of comparing privacy-preserving EEG methods.

Compared with these works, the present paper focuses on a clinical EEG case study in which Gaussian and Laplace perturbation mechanisms are applied to EEG-derived patient-level feature representations and assessed using both statistical utility measures and a downstream machine-learning-based utility check. The emphasis is not on proposing a new DP mechanism or a new EEG classifier, but on examining the practical consequences of subject-level DP perturbation in a multi-hospital EEG processing setting. This focus differs from user-wise perturbation and identity-removal methods because the present work explicitly studies DP noise calibration, patient-level adjacency, sensitivity assumptions, and the resulting privacy--utility trade-off for EEG-derived clinical features.

\section{Proposed Scheme}
\label{sec:proposed-scheme}

A scenario-based privacy-preserving EEG processing scheme is proposed for
multi-hospital clinical neurophysiology settings. The scheme is instantiated
through three deployment scenarios that differ in where anonymization,
aggregation, and model training are performed: client-side anonymization,
centralized server-side anonymization, and decentralized local training. Across
all scenarios, the common objective is to transform raw EEG recordings into
protected EEG-derived feature representations using preprocessing, feature
extraction, and subject-level DP-based perturbation, while limiting the exposure
of raw signals and patient metadata.

\begin{figure}[H]
    \centering
    \includegraphics[width=\textwidth]{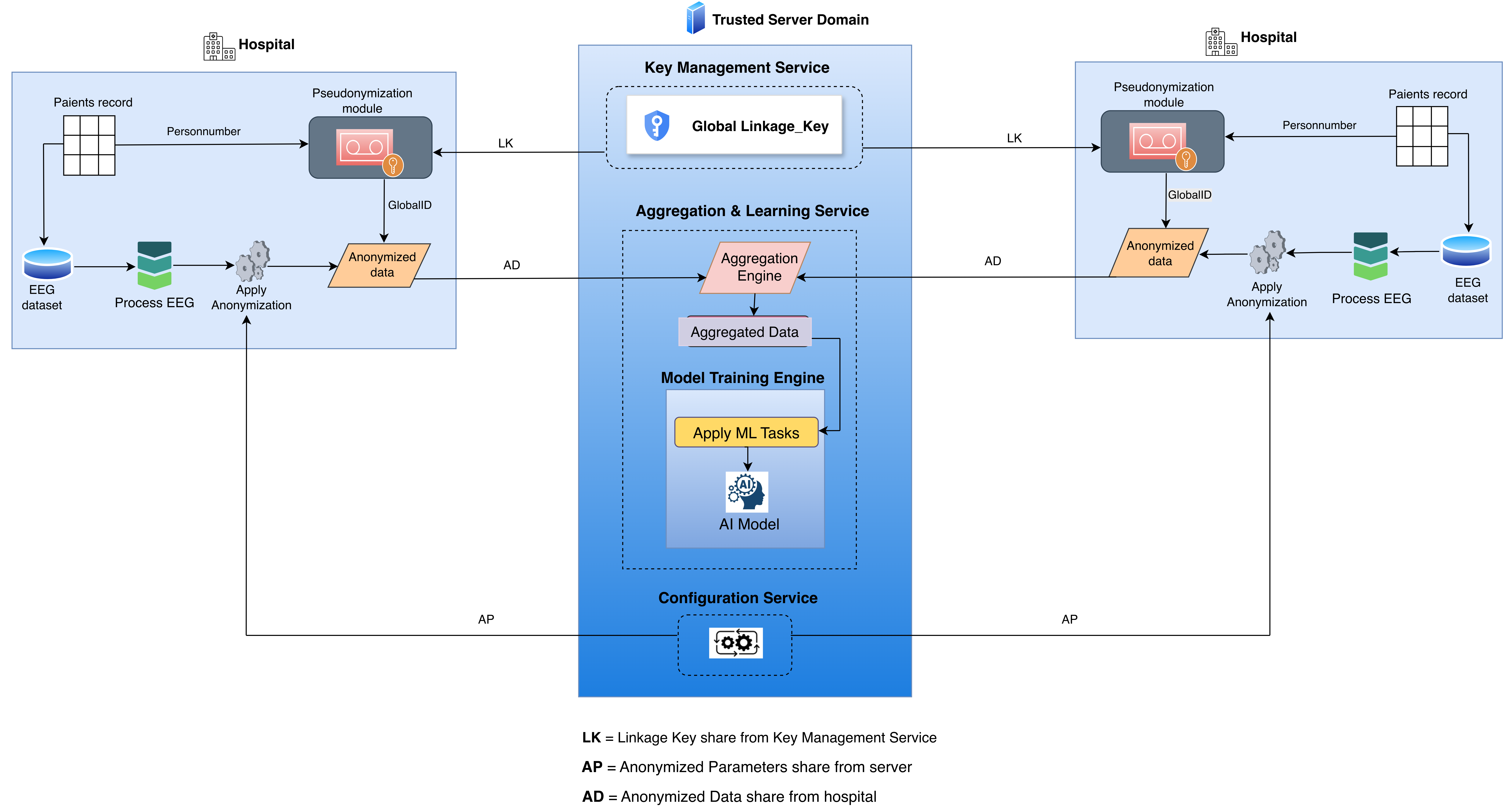}
    \caption{High-level architecture for client-side anonymization (Scenario 1).}
    \label{fig:scenario1}
\end{figure}

\begin{figure}[H]
    \centering
    \includegraphics[width=\textwidth]{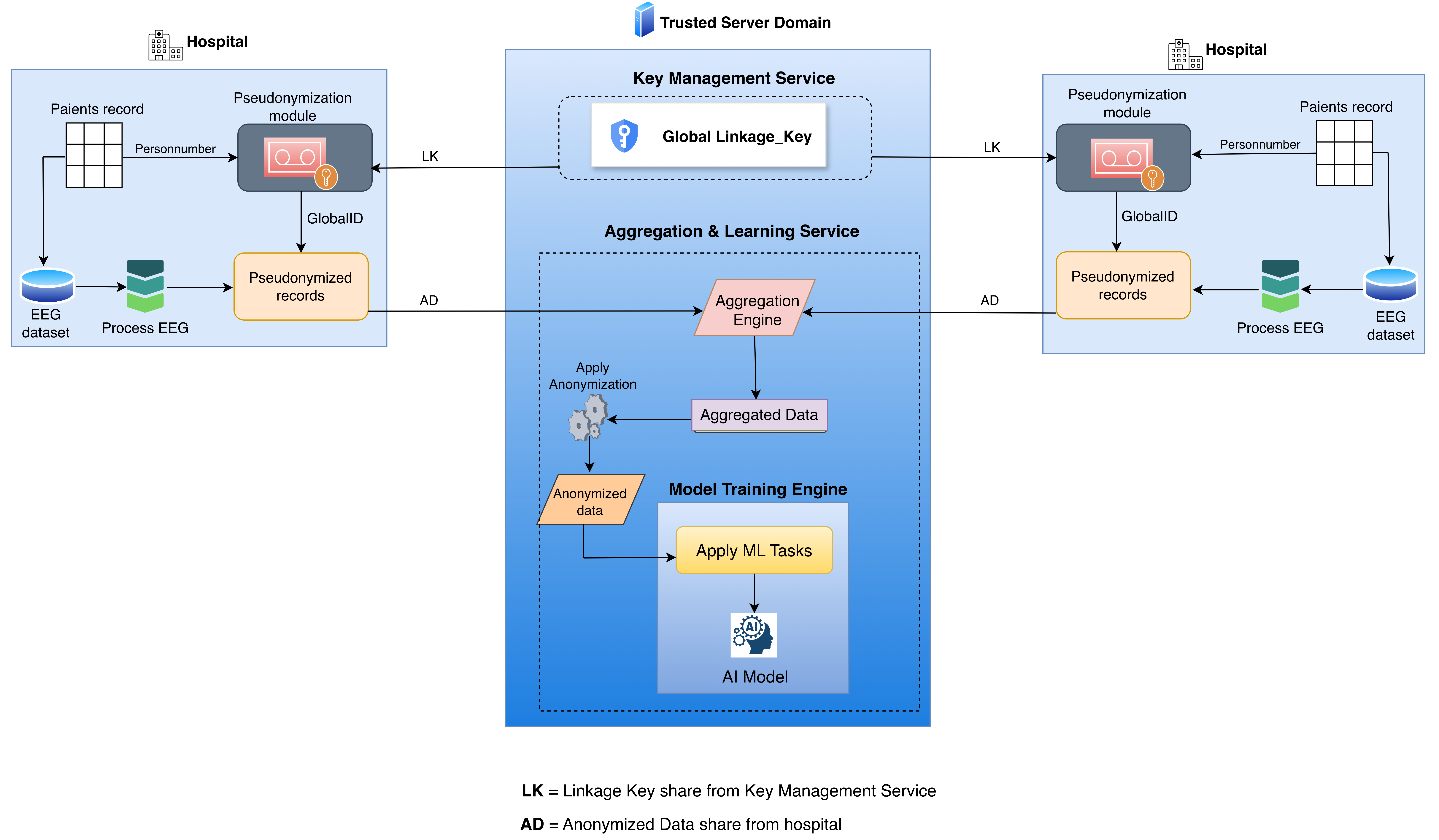}
    \caption{High-level architecture for centralized server-side anonymization (Scenario 2).}
    \label{fig:scenario2}
\end{figure}

\begin{figure}[H]
    \centering
    \includegraphics[width=\textwidth]{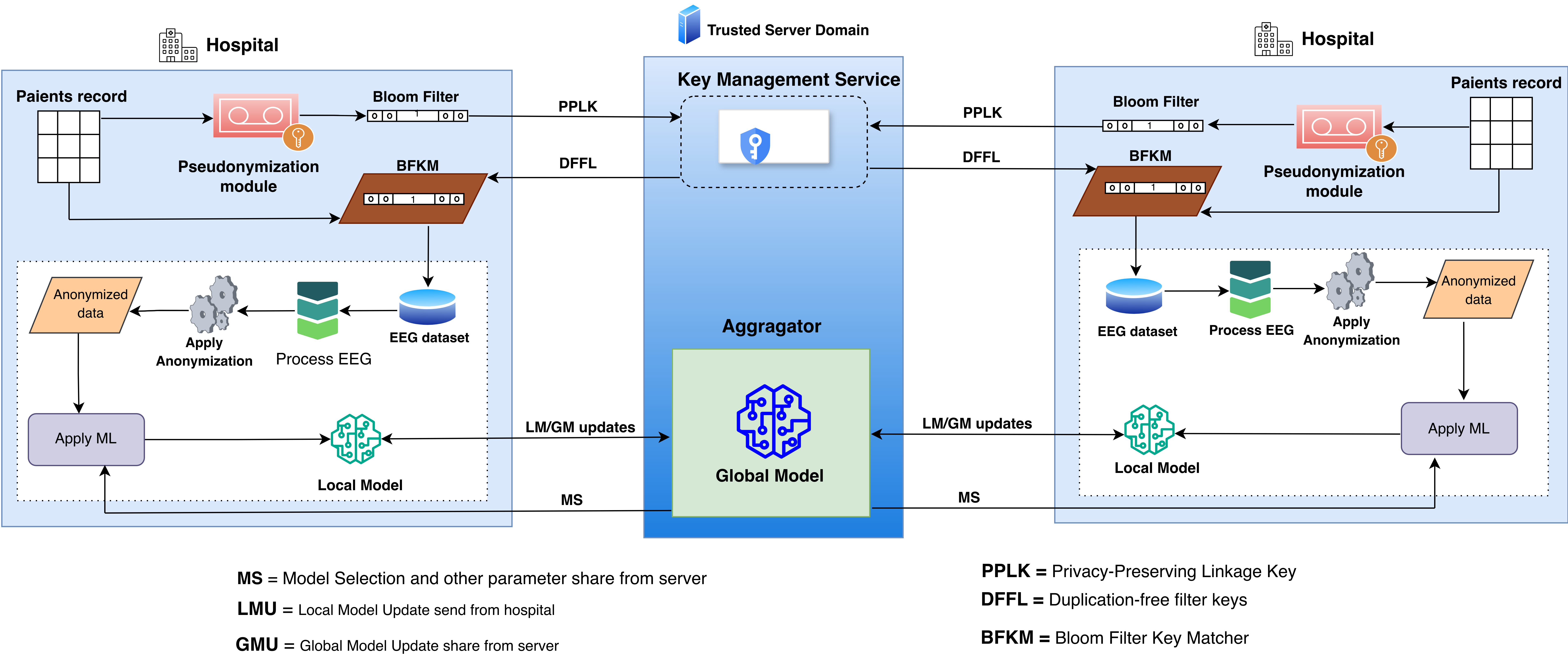}
    \caption{High-level architecture for decentralized local training (Scenario 3).}
    \label{fig:scenario3}
\end{figure}

In Scenario~1, depicted in Figure~\ref{fig:scenario1}, client-side anonymization is performed locally at each hospital before protected feature representations are transferred to the server for aggregation or analysis. In Scenario~2, depicted in Figure~\ref{fig:scenario2}, pseudonymized EEG-derived feature data are  transferred to a trusted server, where server-side anonymization, aggregation, duplicate detection and model training are performed. In Scenario~3, depicted in Figure~\ref{fig:scenario3}, each hospital keeps its EEG data and extracted features locally, applies anonymization within its own environment, and shares only model-related outputs or updates for decentralized learning.

In Scenario~1, the server receives only protected feature representations, while raw EEG data and patient identifiers remain within the local hospital environment. This provides stronger data minimization because the server does not receive identifiable or raw EEG data. A common challenge in multi-hospital settings is that the same patient may appear in more than one hospital. Duplicate patient records can bias aggregation or model training because repeated records give some patients more influence than others. To address this issue, the proposed scheme uses deterministic privacy-preserving linkage. In centralized data-sharing scenarios, a Key Management Service (KMS) distributes a shared linkage key to participating hospitals. Each hospital computes a consistent pseudonym from a canonically encoded shared patient identifier using HMAC-SHA256 \cite{krawczyk1997hmac}:

\begin{equation}
\mathrm{GlobalID}
=
\mathrm{HMAC\text{-}SHA256}_{LK}
\left(\mathrm{Encode}(\mathrm{PatientID})\right),
\end{equation}
where $LK$ is the secret linkage key provided by the KMS and
$\mathrm{Encode}(\cdot)$ denotes an agreed canonical representation of
the shared patient identifier. When participating hospitals possess the
same underlying identifier and apply the same encoding, the same patient
receives the same \texttt{GlobalID} across hospitals, allowing exact
duplicate detection without transmitting the direct identifier.

The \texttt{GlobalID} is used only as an internal linkage identifier for duplicate
detection and record management inside the trusted processing environment. It is not included in the differentially private feature table and is not released to downstream researchers. After duplicate records have been identified and resolved,
the linkage identifier is separated from the DP-protected analytical dataset. This separation is important because releasing stable pseudonyms together with noisy
features could still preserve linkage or membership information across datasets. 
In this paper, pseudonymization refers to replacing direct identifiers with linkage values, whereas DP-based anonymization refers to randomized perturbation of the patient-level feature representation. The HMAC-based \texttt{GlobalID} and Bloom-filter components support record linkage and do not themselves provide differential privacy.

In Scenario~2, anonymization is performed inside the trusted server environment rather than at each hospital. Each participating hospital retains the raw patient identifiers locally, performs basic EEG preprocessing, generates a pseudonymized patient identifier, and transmits pseudonymized EEG-derived feature data to the central server via a secure communication channel. The trusted server then performs aggregation, duplicate detection, DP-based anonymization, and model training.

In Scenario~3, raw EEG data and extracted feature vectors remain within each hospital. Instead of sending patient-level EEG data to the central server, each hospital trains a local model using its own anonymized EEG feature dataset and the privacy/model parameters shared by the central coordinator. The main motivation for this scenario is to reduce direct data sharing between hospitals and the central server. Each hospital performs EEG preprocessing, feature extraction, subject-level anonymization, and local model training within its own secure environment. The server receives only model-related outputs or updates, which can then be aggregated to compute a global model. A common aggregation method is federated averaging \cite{mcmahan2017communication}:
\begin{equation}
w^{(t+1)} =
\sum_{h=1}^{H}
\frac{n_h}{N}
w_h^{(t+1)},
\label{eq:fedavg}
\end{equation}
where $H$ is the number of hospitals, $n_h$ is the number of local training samples at hospital $h$, $N=\sum_{h=1}^{H}n_h$ is the total number of samples, and $w_h^{(t+1)}$ is the locally updated model of hospital $h$ in round $t+1$. 
Differential privacy is applied locally within each hospital before model training. Each hospital preprocesses the raw EEG data, extracts the relevant feature vectors, and applies subject-level DP noise to these extracted features inside its own secure environment. The DP-protected features are then used only for local model training and are not released to the central server or to other hospitals. The central server receives only model-related updates for aggregation. Therefore, the intended privacy mechanism in this scenario is feature-level DP before local training, not DP applied directly to gradients or model updates. This distinction is important because feature-level DP and gradient-level DP require different privacy accounting.

Figure~\ref{fig:scenario3} includes a Bloom-filter-based component for privacy-preserving record linkage across hospitals. This component supports approximate linkage when patient-identifying fields may contain spelling, typographical, or formatting variations that prevent exact matching. In this approach, normalized identifying fields
are divided into character $q$-grams and encoded as Bloom filters, which can then be compared without transmitting the direct identifiers in plaintext \cite{schnell2009pprl}. Bloom-filter encodings, however, do not by themselves provide anonymization and may be vulnerable to frequency and cryptanalytic attacks. A practical deployment therefore requires
keyed hashing, carefully selected encoding and matching parameters, and an explicit threat model \cite{ranbaduge2020securing}. The HMAC construction described above supports exact linkage when hospitals possess the same stable patient identifier, whereas the Bloom-filter component supports approximate linkage when identifying information is not represented identically across hospitals.

\subsection{EEG Processing}
\label{subsec:eeg-processing-proposed}
\begin{figure}[!b]
  \centering
  \includegraphics[width=0.9\textwidth]{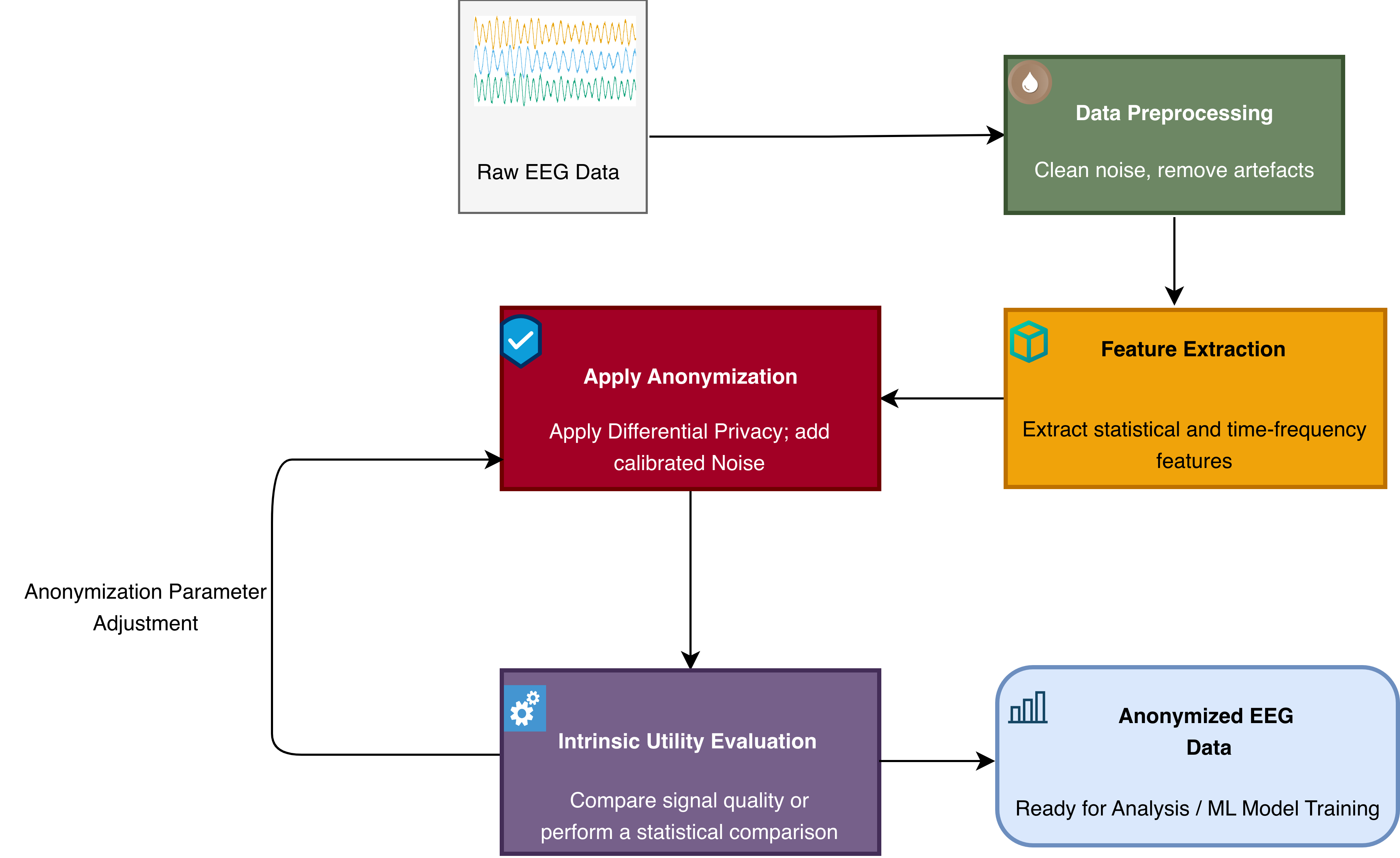}
  \caption{Proposed workflow for EEG processing, DP anonymization, and intrinsic utility assessment in Scenarios 1 and 2.}
  \label{fig:anonymized-without-model-training}
\end{figure}

\begin{figure}[!t]
  \centering
  \includegraphics[width=0.8\textwidth]{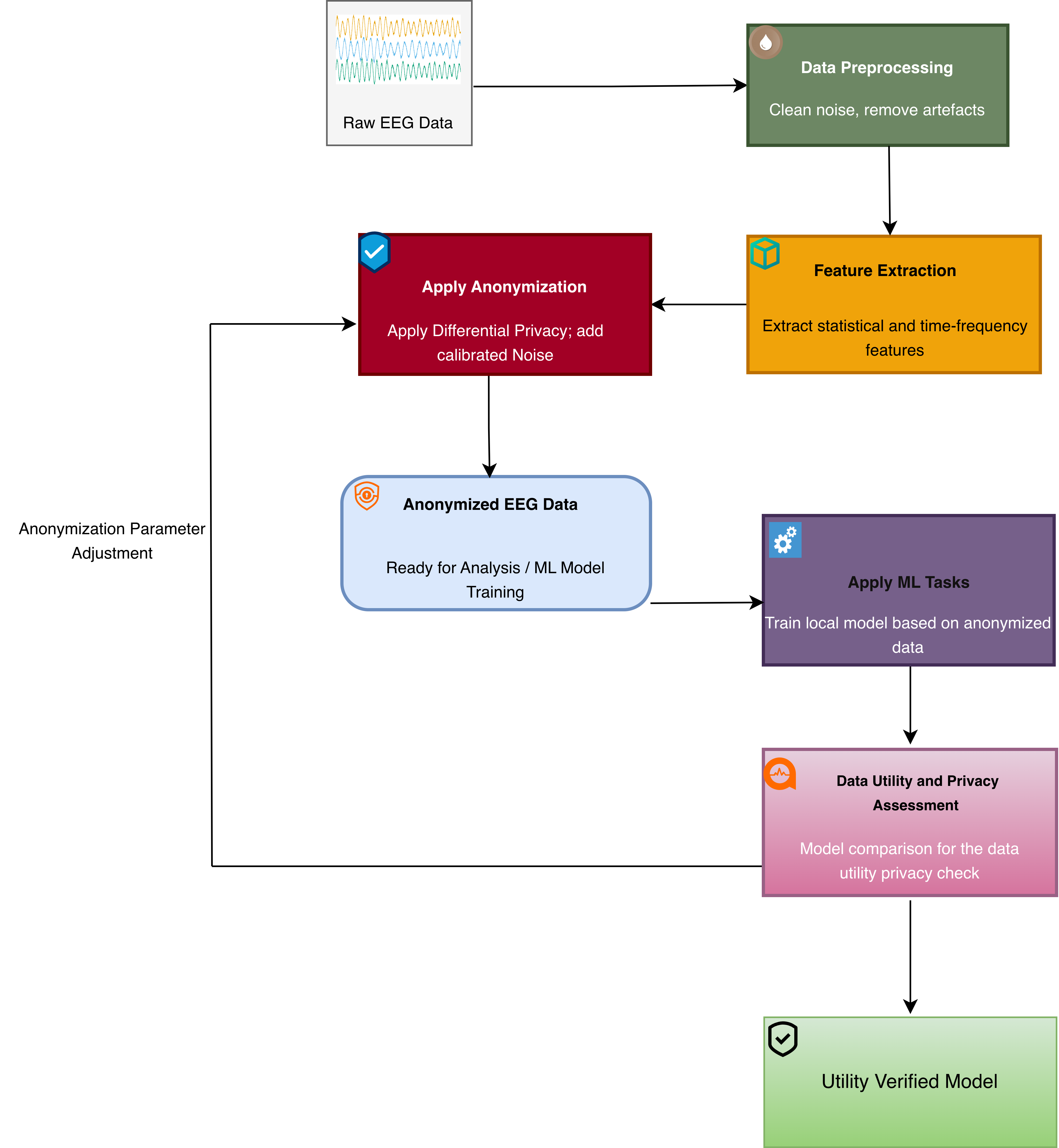}
  \caption{Proposed workflow for EEG processing, DP anonymization, and downstream utility assessment through local model training in Scenario 3.}
  \label{fig:anonymized-with-model-training}
\end{figure}

The proposed scheme uses two EEG processing and anonymization workflows. The first workflow, shown in Figure~\ref{fig:anonymized-without-model-training}, is used in Scenarios~1 and~2, where anonymized EEG-derived features are prepared for data sharing, aggregation, or analysis and are assessed through intrinsic utility measures. The second workflow, shown in Figure~\ref{fig:anonymized-with-model-training}, is used in Scenario~3, where anonymized EEG-derived features remain local and are used for local model training and downstream utility assessment.

The workflow in Figure~\ref{fig:anonymized-without-model-training} consists of four main stages. First, raw EEG data are preprocessed through filtering and artifact reduction to improve signal quality by removing noise such as baseline drift, muscle activity, and eye-blink artifacts. Second, feature extraction converts the cleaned signals into compact statistical and frequency-based representations. Third, DP-based perturbation is applied by adding calibrated noise to the feature vectors to protect patient-level information. Finally, intrinsic utility evaluation compares the anonymized and original features using statistical and signal-based measures to assess whether useful data characteristics are preserved.

\subsubsection{Decentralized EEG Processing and Local Model Training}
\label{subsubsec:decentralized-eeg-processing}
The decentralized workflow extends the same EEG processing and anonymization stages to a local model-training setting. In this scenario, each hospital keeps the EEG data within its local environment. Raw data are preprocessed, features are extracted, and anonymization is applied locally. The anonymized EEG dataset is then used to train a local model. The central server does not receive raw EEG data or subject-level EEG feature data; instead, it receives only trained local model outputs or updates.

The workflow is shown in Figure~\ref{fig:anonymized-with-model-training}. The main processing stages remain preprocessing, feature extraction, and DP-based anonymization. The key difference is the evaluation stage: instead of only checking the intrinsic quality of anonymized features, the decentralized workflow evaluates utility through model performance. A benchmark model is first trained using the original extracted EEG features, and another model is trained using the anonymized EEG features. The performance of these two models is then compared to estimate how much downstream utility is preserved after anonymization.

\subsection{EEG Data Processing Stages}
\label{subsec:eeg-data-processing-stages}

The EEG processing stage prepares raw EEG recordings for anonymization and model training. The input data may come from multiple sources and are assumed to be stored in formats such as \texttt{MAT} or \texttt{HDF5}. Each patient file contains multichannel EEG data arranged as channels by samples. A consistent recording structure is important because it allows the same preprocessing and feature extraction procedure to be applied across patients.

\paragraph{Preprocessing:}
\label{subsubsec:eeg-preprocessing-short} 
EEG preprocessing removes noise and improves signal comparability across patients, channels, and recording sessions. In this work, preprocessing includes band-pass filtering, normalization, and segmentation. The implemented pipeline used a 0.5--70~Hz band-pass filter, a 50~Hz notch filter ($Q=30$), average re-referencing, and 2-s windows with 50\% overlap. Window-level features were averaged separately for each patient to obtain one 363-dimensional patient-level vector. Before clipping and noise addition, the patient-level vectors were stacked into a patient-by-feature matrix and robustly scaled feature-wise using the median and interquartile range calculated across the patients. This normalization was performed after patient-level aggregation.

Band-pass filtering retains relevant EEG frequency components while removing low-frequency drift and high-frequency noise. Segmentation divides continuous EEG recordings into fixed-length epochs for consistent feature extraction. After patient-level aggregation, robust scaling reduces differences in scale among the feature dimensions before clipping and perturbation.

After preprocessing, feature extraction transforms the high-dimensional EEG time series into compact numerical feature vectors. 
In this work, frequency-domain features are derived using Power Spectral Density (PSD), which estimates how signal power is distributed across the main EEG bands: $\delta$, $\theta$, $\alpha$, $\beta$, and $\gamma$. The PSD was estimated using Welch's averaged-periodogram method
\cite{welch1967psd} with a Hann window, a segment length of 256 samples, 50\% overlap (128 samples), and an FFT length of 256 samples.  
These band-power features summarize key neural activity patterns and provide a compact representation of the signals. 
The 363-dimensional feature representation was constructed from 33 EEG channels using 11 per-channel features: line length, root mean square, variance, zero-crossing rate,
$\delta$ band power, $\theta$ band power, $\alpha$ band power, $\beta$ band power, $\gamma$ band power, high-$\gamma$ band power, and the 95\% spectral edge frequency (SEF95). The frequency bands were defined as $\delta=0.5$--$4$~Hz, $\theta=4$--$8$~Hz, $\alpha=8$--$13$~Hz,
$\beta=13$--$30$~Hz, $\gamma=30$--$45$~Hz,
and high-$\gamma=45$--$70$~Hz.
The resulting feature vectors are then used as input to the DP mechanism, allowing anonymization while preserving an interpretable privacy--utility trade-off. DP is applied to EEG-derived features rather than directly to raw EEG time-series signals because feature representations provide a more structured and bounded numerical form for sensitivity control, noise calibration, and utility assessment. 

The utility is assessed in two ways. 
First, intrinsic utility is assessed by comparing the original and
anonymized feature data using root mean square error (RMSE), mean absolute
error (MAE), correlation, and signal-to-noise ratio (SNR). Lower RMSE and MAE, together with higher correlation and SNR, indicate better preservation of the original feature representation. 
Second, in the decentralized model-training workflow, utility is evaluated through downstream model performance by comparing a model trained on original EEG features with a model trained on anonymized EEG features. This combined evaluation helps determine whether the anonymized EEG data remain suitable for later machine-learning tasks.

\subsection{Subject-Level DP}
\label{subsec:subject-level-dp}

\begin{figure}[!t]
  \centering
  \includegraphics[width=\textwidth]{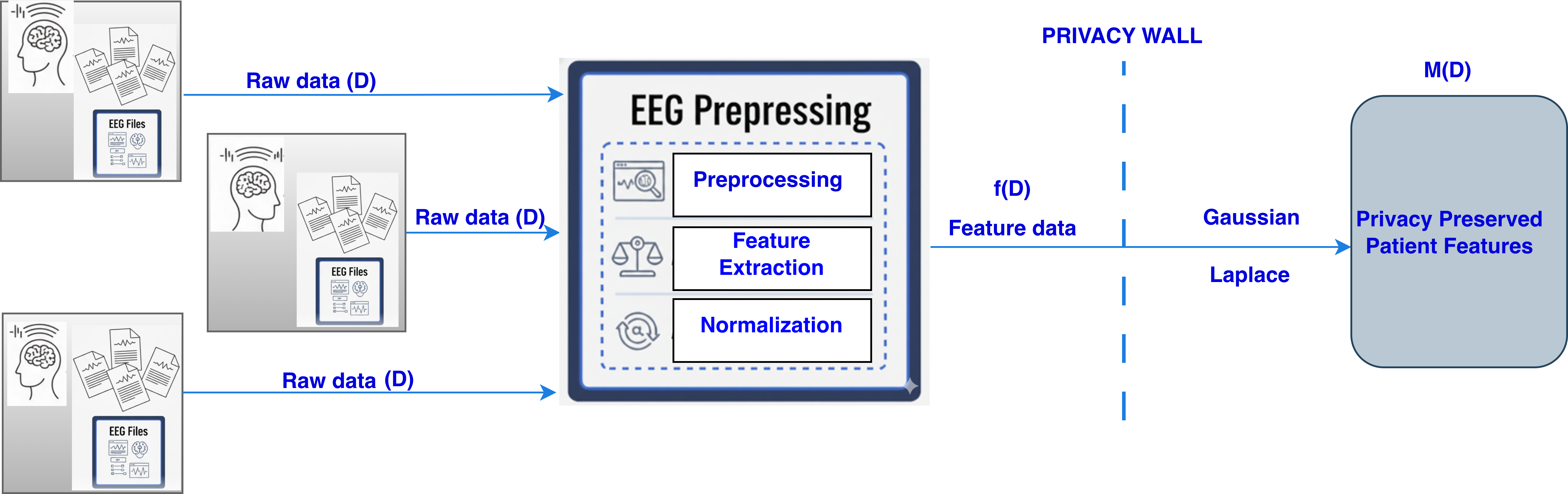}
  \caption{Subject-level anonymization of patient EEG feature data.}
  \label{fig:subject-level-anonymization}
\end{figure}

The anonymization stage applies subject-level DP to the extracted EEG feature vectors. The privacy setting is limited to subject-level DP, where the contribution of one patient is protected according to the selected adjacency relation and privacy parameters. Subject-level privacy is selected because the goal is to protect the contribution of an entire patient rather than a single EEG epoch or segment. Figure~\ref{fig:subject-level-anonymization} illustrates the application of subject-level DP to patient EEG feature data.

For the formal description, let $D=(x_1,\ldots,x_n)$ denote a fixed-size table of patient-level EEG feature vectors, where each $x_i\in\mathbb{R}^d$ represents the features extracted for one patient. This paper uses a bounded, replace-one subject-level adjacency relation: two datasets $D$ and $D'$ are neighboring if they contain the same number of patients and differ in the feature vector of exactly one patient. 
In this implementation, the privacy unit is one patient. The pipeline constructs one aggregated EEG feature vector for each patient after preprocessing and feature extraction. Differential privacy is therefore applied to patient-level feature vectors, not to individual EEG files, epochs, segments, or sliding windows. If a patient has multiple recordings or epochs, these are first summarized into a single patient-level representation before the DP mechanism is applied. Consequently, the neighboring datasets in this paper differ by the complete feature contribution of one patient, which is consistent with the bounded, replace-one subject-level adjacency relation used in the formal definition.

A randomized mechanism $M$ satisfies $(\varepsilon,\delta)$-differential privacy under this adjacency relation if, for all neighboring datasets $D,D'$ and all measurable output sets $S$,
\begin{equation}
\Pr[M(D)\in S]
\leq
e^{\varepsilon}\Pr[M(D')\in S]+\delta .
\label{eq:dp-definition-proposed}
\end{equation}
The parameter $\varepsilon$ controls the privacy loss, while $\delta$ denotes a small probability of failure. Smaller values of $\varepsilon$ provide stronger privacy but require more noise, which can reduce feature utility.

The mechanism considered in this paper operates on a patient-level EEG feature table. In the implementation, the privacy unit is one patient, and each patient is represented by one feature vector after preprocessing and feature extraction. The resulting feature dimension is $d = 363$. 
The implementation follows a bounded, replace-one subject-level adjacency relation, where two neighboring datasets contain the same number of patients and differ in the complete feature contribution of exactly one patient.

Before noise addition, the patient-level vectors were assembled into a matrix with 363 feature columns. Each value was centered by the corresponding feature-wise median and divided by the corresponding interquartile range, both calculated across patients. Interquartile ranges less than or equal to $10^{-9}$ were replaced with $1.0$ before division.  
The scaling statistics were fitted within each invocation of the perturbation routine and remained unchanged during its clipping and noise-addition steps. Because these statistics were calculated from the private patient dataset, the sensitivity bounds below assume a fixed scaling transformation and do not establish an end-to-end privacy guarantee for the complete preprocessing pipeline.

Each scaled vector is then clipped to a fixed $L_2$-norm threshold $C$:
\begin{equation}
\bar{x}_i =
x_i \cdot
\min\left(1,\frac{C}{\|x_i\|_2}\right).
\label{eq:clipping}
\end{equation}
In the implementation, the clipping threshold is $C = 2.0$. 

The clipped table is denoted by $\bar{D}=(\bar{x}_1,\ldots,\bar{x}_n)$. Conditional on fixed scaling parameters, and under the replace-one adjacency relation, the $L_2$ sensitivity of the full clipped table, viewed as one concatenated vector, is bounded by
\begin{equation}
\Delta_2 \leq 2C .
\label{eq:l2-sensitivity}
\end{equation}
With $C=2.0$, this gives
\begin{equation}
\Delta_2 = 4.0 .
\label{eq:l2-sensitivity-value}
\end{equation}

For coordinate-wise Laplace noise applied to the full feature vector, the corresponding $L_1$ sensitivity must be considered. If only $L_2$ clipping is applied and the scaling parameters are treated as fixed, the $L_1$ sensitivity is bounded by
\begin{equation}
\Delta_1 \leq 2\sqrt{d}C ,
\label{eq:l1-sensitivity}
\end{equation}
where $d$ is the feature dimension. Since the implementation uses $d=363$ and $C=2.0$, the resulting bound is
\begin{equation}
\Delta_1 \leq 2\sqrt{363}\cdot 2.0 \approx 76.21 .
\label{eq:l1-sensitivity-value}
\end{equation}

\begin{table}[!t]
\centering
\caption{Sensitivity bounds and perturbation parameters used in the patient-level EEG feature experiments.}
\begin{tabular}{|p{0.5\linewidth}|c|}
\hline
\textbf{Component} & 
\textbf{Implemented value or derived bound} \\ \hline
Privacy unit & Single patient \\ \hline

Feature dimension & $d=363$ \\ \hline

Clipping threshold & $C=2.0$ \\ \hline
$L_2$ sensitivity bound & $\Delta_2 = 2C = 4.0$ \\ \hline
$L_1$ sensitivity bound & $\Delta_1 \leq 2\sqrt{d}C \approx 76.21$ \\ \hline
Gaussian $\delta$ & $\delta = 10^{-5}$ \\ \hline
Tested $\varepsilon$ values & $\{0.5,1.0,2.0,5.0,10.0,20.0\}$ \\ \hline
Gaussian sensitivity used & $4.0$ \\ \hline
Laplace sensitivity used in implementation & $4.0$ \\ \hline
\end{tabular}
\label{tab:dp-calibration-config}
\end{table}

Table~\ref{tab:dp-calibration-config} summarizes the differential privacy configuration used in the implementation. The Gaussian and Laplace mechanisms considered in this study add noise to the clipped patient-level feature vectors. For the Gaussian mechanism, noise is sampled from $\mathcal{N}(0,\sigma^2 I_d)$ and added to each clipped vector. The implementation uses the analytic Gaussian mechanism with $\Delta_2=4.0$ and $\delta=10^{-5}$. The Gaussian noise scale is therefore calibrated using the $L_2$ sensitivity of the clipped patient-level vector and the selected privacy budget $\varepsilon$.

For the Gaussian mechanism, the classical sufficient calibration is
\begin{equation}
\sigma \geq
\frac{\Delta_2\sqrt{2\ln(1.25/\delta)}}{\varepsilon}.
\label{eq:classical-gaussian}
\end{equation}
However, because this classical expression is commonly stated under restrictive parameter conditions, the implementation uses the analytic Gaussian mechanism. The Gaussian experiments are reported for
\begin{equation}
\varepsilon \in \{0.5,1.0,2.0,5.0,10.0,20.0\},
\qquad
\delta = 10^{-5},
\qquad
\Delta_2 = 4.0 .
\label{eq:gaussian-configuration}
\end{equation}

For the Laplace mechanism, independent coordinate-wise noise is added to each clipped feature coordinate. The implementation used a sensitivity value of $4.0$, corresponding to $2C$. Therefore, the Laplace scale used in the experiments was
\begin{equation}
b_{\mathrm{impl}} =
\frac{4.0}{\varepsilon}.
\label{eq:laplace-implemented-scale}
\end{equation}
This gives the following implemented noise scales:
\begin{equation}
b_{\mathrm{impl}} \in
\{8.0,4.0,2.0,0.8,0.4,0.2\}
\quad
\text{for}
\quad
\varepsilon \in \{0.5,1.0,2.0,5.0,10.0,20.0\}.
\label{eq:laplace-implemented-scales}
\end{equation}

For a formal coordinate-wise Laplace release of the complete 363-dimensional patient-level vector under replace-one subject-level adjacency, the required $L_1$ sensitivity bound is instead
\begin{equation}
\Delta_1 \leq 2\sqrt{363}\cdot 2.0 \approx 76.21
\label{eq:laplace-formal-l1}
\end{equation}

The corresponding formally calibrated Laplace scale would be
\begin{equation}
b_{\mathrm{formal}} =
\frac{\Delta_1}{\varepsilon}
=
\frac{76.21}{\varepsilon}
\label{eq:laplace-formal-scale}
\end{equation}

Thus, the formally calibrated Laplace scales would be approximately 
\begin{equation}
b_{\mathrm{formal}} \in
\{152.42,76.21,38.10,15.24,7.62,3.81\}
\quad
\text{for}
\quad
\varepsilon \in \{0.5,1.0,2.0,5.0,10.0,20.0\}.
\label{eq:laplace-formal-scales}
\end{equation}

Consequently, the Gaussian experiments use the implemented $L_2$ sensitivity bound $\Delta_2=4.0$, whereas the Laplace implementation uses $b_{\mathrm{impl}}=4.0/\varepsilon$ rather than the full-vector scale $b_{\mathrm{formal}}=76.21/\varepsilon$. The Laplace results therefore characterize utility under the implemented perturbation magnitudes and are not claimed to provide formal full-vector $\varepsilon$-DP guarantees under replace-one subject-level adjacency. This comparison remains informative
for showing how perturbation magnitude affects EEG feature utility, while Equations~\ref{eq:laplace-formal-l1}--\ref{eq:laplace-formal-scales} give the calibration required for a formal full-vector Laplace release.
The Gaussian mechanism is calibrated using the analytic Gaussian accountant \cite{balle2018improving}, including for the larger $\varepsilon$ values evaluated in this study.

\subsubsection{Applying anonymization}
The implemented anonymization procedure follows the same high-level structure for both perturbation mechanisms. Each patient-level EEG feature vector is first scaled and clipped to bound its contribution. For each mechanism, a noise scale is determined from the selected parameters and sensitivity value, and random noise is then added to produce the perturbed feature vector. The resulting privatized feature table is used for intrinsic utility analysis or downstream model training. The clipping step is essential because it limits the maximum influence that any single patient can have on the released feature table.

The Gaussian mechanism uses the $L_2$ sensitivity of the clipped patient-level feature vector. In contrast, a formally calibrated coordinate-wise Laplace mechanism for the full released feature vector requires the corresponding $L_1$ sensitivity, with scale $b = \Delta_1 / \epsilon$. The Laplace experiments used the implemented sensitivity value $4.0$ and scale $b_{\mathrm{impl}}=4.0/\epsilon$. Accordingly, these results evaluate the implemented Laplace perturbation settings and are not claimed to provide a formally calibrated full-vector Laplace-DP guarantee under replace-one subject-level adjacency. The original raw EEG recordings and direct patient identifiers are not included in the released feature table.

\begin{algorithm}[!b] 
\caption{Gaussian perturbation of clipped EEG feature vectors} \label{alg:Gaussian-anonymization} 
\footnotesize \begin{algorithmic}[1] 
\Function{Gaussian-Anonymization}{$D$, $C$, $\varepsilon$, $\delta$, accountant} 
\Statex \textit{Step 1: Clip patient-level feature vectors} 
\For{each patient vector $x_i \in D$} \State $\bar{x}_i \gets x_i \cdot \min\left(1, C/\|x_i\|_2\right)$ \EndFor \Statex \textit{Step 2: Calibrate noise to subject-level sensitivity} \State $\Delta_2 \gets 2C$ \Comment{replace-one adjacency} \State $\sigma \gets \textsc{GaussianScale}(\Delta_2,\varepsilon,\delta,\text{accountant})$ \Statex \textit{Step 3: Add Gaussian noise} 
\For{each clipped vector $\bar{x}_i$} 
\State Draw $z_i \sim \mathcal{N}(0,\sigma^2 I_d)$ \State $\tilde{x}_i \gets \bar{x}_i + z_i$ \State Add $\tilde{x}_i$ to $\widetilde{D}$ 
\EndFor \State \Return $\widetilde{D}$ 
\EndFunction \end{algorithmic} 
\end{algorithm} 

\begin{algorithm}[!t] 
\caption{Laplace perturbation of clipped EEG feature vectors} \label{alg:laplace-anonymization} \footnotesize \begin{algorithmic}[1] \Function{Laplace-Anonymization}{$D$, $C$, $\varepsilon$, $d$} \Statex \textit{Step 1: Clip patient-level feature vectors} \For{each patient vector $x_i \in D$} \State $\bar{x}_i \gets x_i \cdot \min\left(1, C/\|x_i\|_2\right)$ \EndFor \Statex \textit{Step 2: Calibrate coordinate-wise Laplace noise} \State $\Delta_1 \gets 2\sqrt{d}C$ \Comment{replace-one adjacency and $L_2$ clipping} \State $b \gets \Delta_1/\varepsilon$ \Statex \textit{Step 3: Add Laplace noise} \For{each clipped vector $\bar{x}_i$} \State Draw $z_i \in \mathbb{R}^d$, where each $z_{ij} \sim \mathrm{Lap}(0,b)$ \State $\tilde{x}_i \gets \bar{x}_i + z_i$ \State Add $\tilde{x}_i$ to $\widetilde{D}$ \EndFor \State \Return $\widetilde{D}$ \EndFunction \end{algorithmic} \end{algorithm}

Algorithms~\ref{alg:Gaussian-anonymization} and \ref{alg:laplace-anonymization} summarize the two perturbation procedures considered. In both algorithms, the input feature vectors are assumed to be the patient-level feature vectors after the scaling step described above. Algorithm~\ref{alg:Gaussian-anonymization} describes the Gaussian perturbation procedure, where the noise scale is obtained from the selected Gaussian accountant. Algorithm~\ref{alg:laplace-anonymization} describes the formally calibrated coordinate-wise Laplace procedure based on the $L_1$ sensitivity bound.

\section{Performance Analysis}
\label{chap:performance} 
The proposed privacy-preserving EEG framework was implemented in Python using MNE-Python for EEG preprocessing \cite{gramfort2013mne}, NumPy and SciPy for feature extraction, scikit-learn for downstream classification \cite{pedregosa2011scikit}, and IBM DiffPrivLib for implementing the Gaussian and Laplace perturbation mechanisms. Experiments were conducted in Docker-based local environments, with Jupyter Notebook used for development and testing. A clinical EEG dataset was used, containing 122 EEG recording files, 8 GB of recordings, around 33 channels, and sampling rates of 250--500 Hz.

The implementation constructs one patient-level EEG feature vector per patient. Raw EEG files are first grouped by patient identifier, preprocessed, segmented into 2-second windows with 50\% overlap, and converted into statistical and frequency-domain EEG features. Window-level features belonging to the same patient are then aggregated by the mean to obtain a single patient-level representation. Labels are assigned at the patient-level from the available metadata files using strict seizure-related keywords.

Table~\ref{tab:reproducibility_config} summarizes the main dataset, preprocessing, privacy, and model-evaluation settings used in the experiments. Each privacy configuration was evaluated using one noise realization. The resulting measurements are reported as a case-study comparison of utility trends across the tested privacy budgets. 

The analysis focuses on three aspects: statistical utility preservation, frequency-domain signal distortion, and downstream machine learning performance. The objective is to assess the privacy--utility trade-off introduced by the Gaussian and Laplace perturbations and to determine how anonymization affects both signal quality and classification performance. The evaluation uses both statistical and classification-based metrics. Statistical metrics compare the original and anonymized EEG features without training a machine learning model, measuring the extent to which anonymization preserves the structure and distribution of the original data. Classification metrics are then used to evaluate whether anonymized EEG features remain useful for downstream epilepsy detection.

\begin{table}[H]
\centering
\caption{Reproducibility configuration for the patient-level EEG anonymization experiments.
}
\begin{tabular}{|p{0.35\linewidth}|p{0.6\linewidth}|}
\hline
\textbf{Component} & \textbf{Value used in implementation} \\ \hline
Programming environment & Python 3.13.15, Jupyter Notebook 7.2.2\\ \hline
Main libraries & MNE-Python 1.10.1, NumPy 2.1.3, SciPy 1.16.3, scikit-learn 1.6.1, IBM DiffPrivLib 0.6.6 \\ \hline
Dataset source & Anonymized clinical EEG recordings made available through the VIKING project \textsuperscript{*} \footnotemark 
\\ \hline
Dataset size & 122 EEG recording files from 17 different patients, around 8 GB of recordings \\ \hline
EEG channels and sampling rate & Around 33 channels; sampling rates of 250--500 Hz \\ \hline
Scaling before perturbation & Feature-wise robust scaling of the aggregated patient vectors using the median and interquartile range, followed by $L_2$ clipping \\ \hline
Patient-level aggregation & Mean aggregation of window-level features per patient \\ \hline
Feature dimension & $d=363$ \\ \hline
Clipping threshold & $C=2.0$ \\ \hline
Adjacency relation & Bounded replace-one subject-level adjacency \\ \hline
Privacy unit & One patient \\ \hline

$\sigma$ values & Obtained directly from the IBM DiffPrivLib \texttt{GaussianAnalytic} mechanism using $\Delta_2=4.0$ and $\delta=10^{-5}$ for each tested $\epsilon$. \\ \hline

Gaussian $L_2$ sensitivity used in the experiments & $\Delta_2 = 2C = 4.0$ \\ \hline
General $L_1$ sensitivity bound for coordinate-wise Laplace release & $\Delta_1 \leq 2\sqrt{d}C \approx 76.21$ \\ \hline

Gaussian $\delta$ & $10^{-5}$ \\ \hline
Laplace mechanism implementation & Coordinate-wise Laplace noise generated using IBM DiffPrivLib \\ \hline
Laplace sensitivity used in implementation & $4.0$ \\ \hline
Tested $\varepsilon$ values & $\{0.5,1.0,2.0,5.0,10.0,20.0\}$ \\ \hline
Implemented Laplace scales & $\{8.0,4.0,2.0,0.8,0.4,0.2\}$ for $\varepsilon=\{0.5,1.0,2.0,5.0,10.0,20.0\}$ \\ \hline
Formal Laplace scales under full $L_1$ calibration & $\{152.42,76.21,38.10,15.24,7.62,3.81\}$ for $\varepsilon=\{0.5,1.0,2.0,5.0,10.0,20.0\}$ \\ \hline

Noise repetitions & One noise realization per $\varepsilon$ setting 
\\ \hline

Threshold-selection status & Cohort-level threshold optimization using the aggregated LOSO predictions \\ \hline

Threshold grid & Candidate thresholds considered in the experiments: 0.20--0.80 for the baseline evaluation and 0.05--0.95 for the DP evaluations. \\ \hline
\end{tabular}
\label{tab:reproducibility_config}
\end{table}
\footnotetext{\textsuperscript{* } \texttt{https://viking-project.org}}

\subsection{Privacy--Utility Assessment of the Gaussian Mechanism} 

The Gaussian mechanism adds noise sampled from a Gaussian distribution, where the noise level is controlled by the privacy budget $\epsilon$, the privacy failure probability $\delta$, and the $L_2$ sensitivity of the released feature vector. In this implementation, Gaussian noise was calibrated using IBM DiffPrivLib's \texttt{GaussianAnalytic} mechanism with $\Delta_2=4.0$ and $\delta=10^{-5}$. 
The implementation used analytic Gaussian noise calibration with $\Delta_2 = 4.0$ for the clipped feature representation. This sensitivity bound assumes a fixed scaling transformation; it does not account for changes in the median and interquartile range when a patient is replaced in the underlying dataset. 
The actual Gaussian noise scale $\sigma$ used for each privacy budget is reported in Table~\ref{tab:gaussian_sigma_values}. 
The evaluation then examines how these calibrated $\epsilon$ values affect distortion, correlation, and signal-to-noise ratio.

\begin{table}[!t]
\centering
\caption{Gaussian-DP calibration across $\epsilon$ values.}
\begin{tabular}{|c|c|c|c|}
\hline
\textbf{$\epsilon$} & \textbf{$\delta$} & \textbf{$\Delta_2$} & \textbf{Noise scale $\sigma$} \\ \hline
0.5  & $10^{-5}$ & 4.0 & 28.1273 \\ \hline
1.0  & $10^{-5}$ & 4.0 & 14.9225 \\ \hline
2.0  & $10^{-5}$ & 4.0 & 7.9752  \\ \hline
5.0  & $10^{-5}$ & 4.0 & 3.5675  \\ \hline
10.0 & $10^{-5}$ & 4.0 & 1.9996  \\ \hline
20.0 & $10^{-5}$ & 4.0 & 1.1602  \\ \hline
\end{tabular}
\label{tab:gaussian_sigma_values}
\vspace{2pt}
\end{table}

\begin{table}[!t]
\centering
\caption{Privacy--utility trend across $\epsilon$ values for the analytically calibrated Gaussian mechanism.}
\begin{tabular}{|c|c|c|c|c|}
\hline
\textbf{$\epsilon$} & \textbf{RMSE} & \textbf{MAE} & \textbf{Correlation} & \textbf{SNR (dB)} \\ \hline
0.5  & 28.7048 & 22.9830 & -0.0090 & -48.7376 \\ \hline
1.0  & 14.7640 & 11.7113 & -0.0052 & -42.9626 \\ \hline
2.0  & 7.9066  & 6.3135  & 0.0105  & -37.5382 \\ \hline
5.0  & 3.5468  & 2.8286  & 0.0220  & -30.5752 \\ \hline
10.0 & 1.9988  & 1.5913  & 0.0522  & -25.5937 \\ \hline
20.0 & 1.1561  & 0.9198  & 0.0743  & -20.8387 \\ \hline
\end{tabular}
\label{tab:gaussian_epsilon_results}
\vspace{2pt}
\end{table}

\begin{figure}[!t]
  \centering
\includegraphics[width=0.6\textwidth]{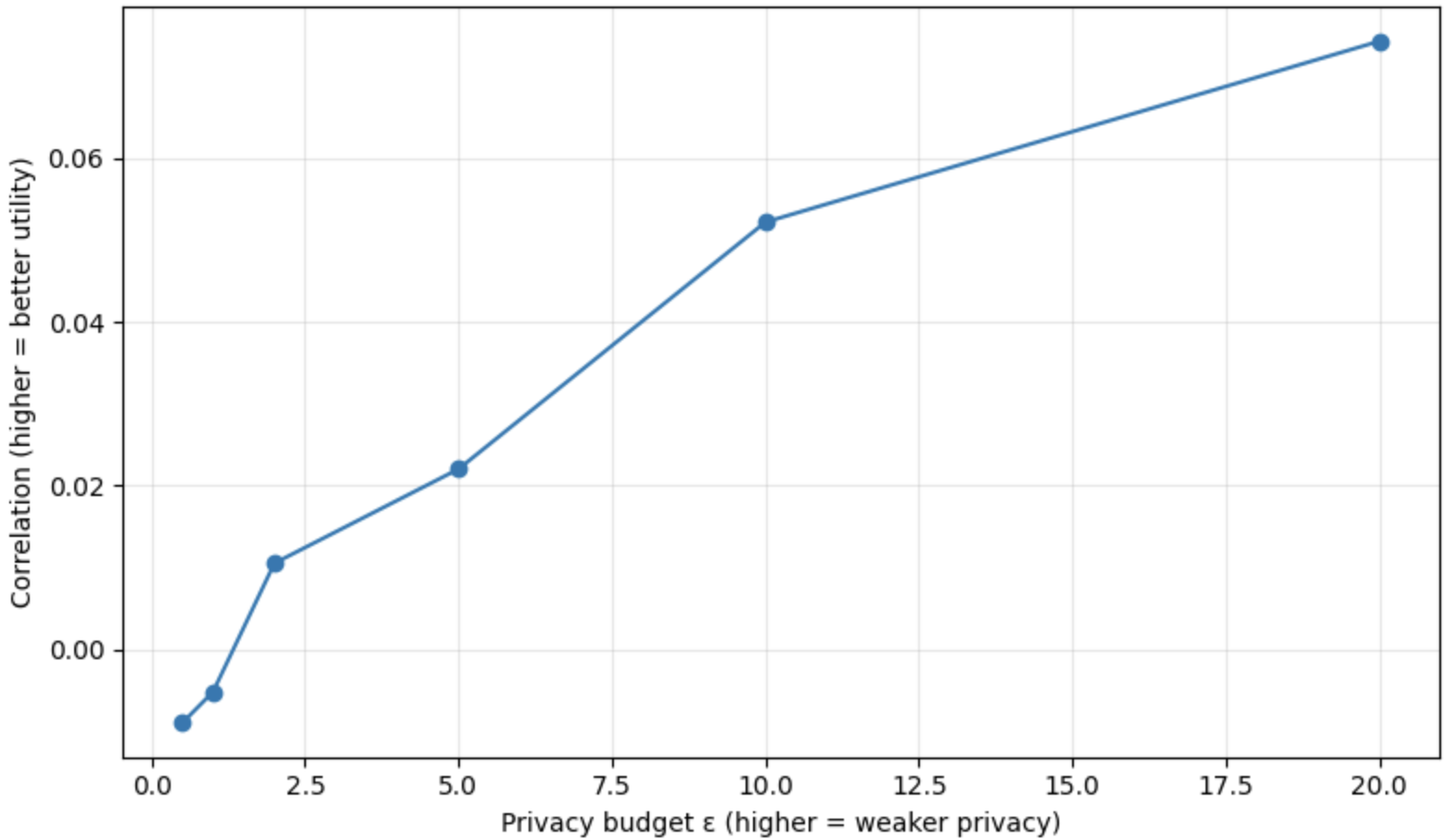}
  \caption{Utility trend under the analytically calibrated Gaussian mechanism.}
  \label{fig:utility-loss-gaussian}
  \vspace{-5pt}
\end{figure}

As shown in Table~\ref{tab:gaussian_epsilon_results} and
Figure~\ref{fig:utility-loss-gaussian}, increasing $\epsilon$ reduces RMSE and MAE, indicating that the perturbed feature representation becomes closer to the original feature representation. The correlation and SNR also improve as $\epsilon$ increases. However, the correlation remains low across all tested values and the SNR remains negative, indicating that the added noise remains large relative to the evaluated feature representation. This indicates a substantial utility loss under the Gaussian mechanism, especially in stronger privacy settings. 

The results indicate that preserving EEG feature utility with the Gaussian mechanism is difficult in this setting. Even at larger privacy budgets, the correlation remains low and the SNR remains negative. This suggests that the tested Gaussian configurations introduce substantial distortion even under analytic Gaussian calibration.

\subsection{Privacy--Utility Assessment of the Laplace Mechanism} 
Coordinate-wise Laplace noise is formally calibrated using the $L_1$ sensitivity of the released object and the privacy parameter $\epsilon$. Here, the released object is the complete patient-level EEG feature vector under replace-one subject-level adjacency. With $L_2$ clipping, the corresponding bound is $\Delta_1\leq2\sqrt{d}C$; for $d=363$ and $C=2.0$, this gives $\Delta_1\leq76.21$. The implementation instead used the sensitivity value $4.0$ and scale $b_{\mathrm{impl}}=4.0/\epsilon$. The resulting experiments quantify utility under these implemented Laplace perturbation magnitudes, while the scale required for a formal full-vector Laplace release, $b_{\mathrm{formal}}=76.21/\epsilon$, is derived separately above.

\begin{table}[!b]
\centering
\caption{Utility trend under the implemented Laplace scales $b_{\mathrm{impl}}=4.0/\epsilon$.}
\begin{tabular}{|c|c|c|p{11.6cm}|}
\hline
\textbf{$\epsilon$} & \textbf{Corr.} & \textbf{SNR (dB)} & \textbf{Interpretation} \\
\hline
0.5  & 0.019 & -40.79 & Largest perturbation setting among the tested values, with very low utility. \\
1.0  & 0.024 & -34.51 & Large perturbation setting; utility remains low. \\
2.0  & 0.043 & -28.66 & 
Large perturbation setting; correlation and SNR begin to improve. \\
5.0  & 0.103 & -20.54 & Moderate perturbation setting; utility improves but remains limited. \\
10.0 & 0.201 & -14.80 & Lower perturbation setting; utility improves compared with smaller $\epsilon$ values. \\
20.0 & 0.335 &  -8.61 & Smallest perturbation setting among the tested values, with the highest observed utility. \\
\hline
\end{tabular}
\label{tab:laplace_epsilon_choice}
\end{table}

\begin{figure}[!b]
  \centering
\includegraphics[width=0.6\textwidth]{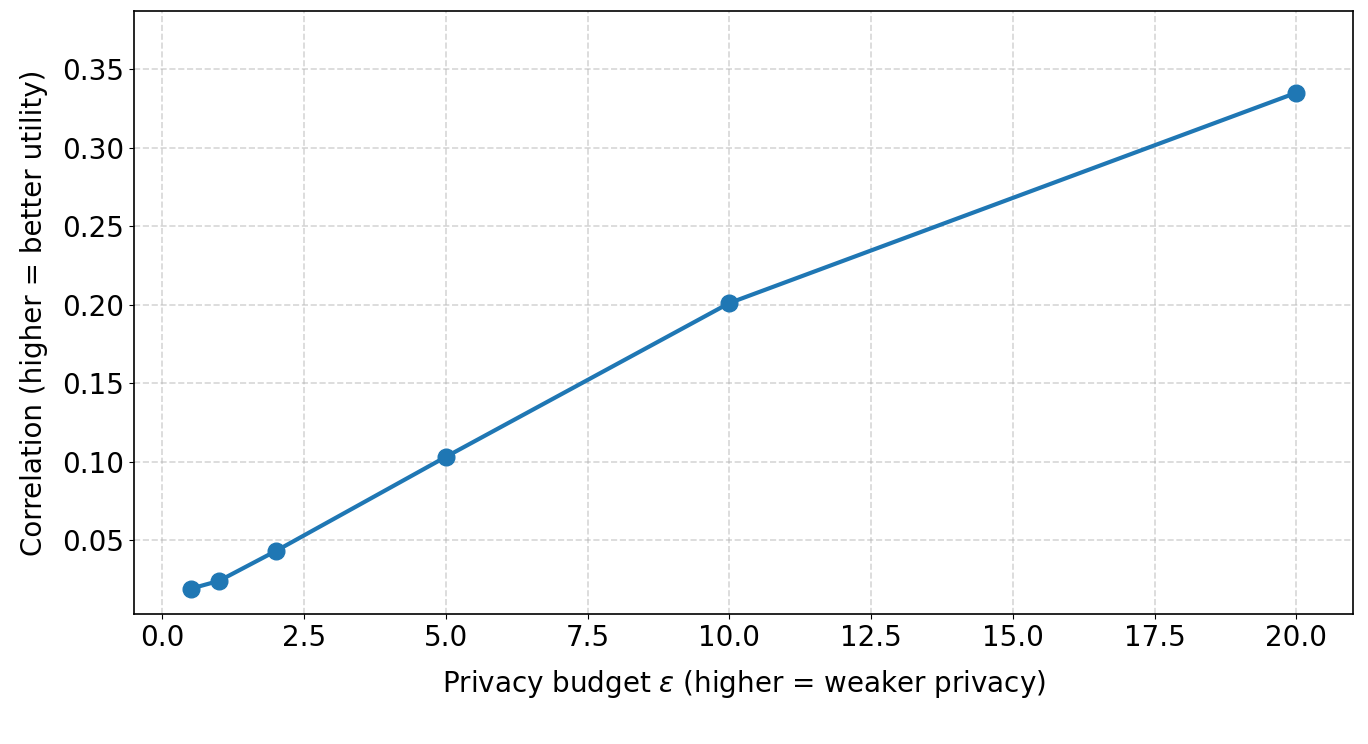}
\vspace{-10pt}
  \caption{Correlation between the original and perturbed feature values under the implemented Laplace perturbation with $b_{\mathrm{impl}}=4.0/\epsilon$.
  }  
  \label{fig:privacy-laplace}
\end{figure}

Table~\ref{tab:laplace_epsilon_choice} shows that increasing $\epsilon$ reduces the Laplace perturbation scale used in the experiments and improves both correlation and SNR, while
Figure~\ref{fig:privacy-laplace} visualizes the corresponding correlation trend. Smaller $\epsilon$ values correspond to larger perturbation magnitudes and lower utility, while larger $\epsilon$ values preserve more feature structure and correspond to smaller perturbation magnitudes in the implemented sweep. Because the Laplace implementation used the sensitivity value $4.0$ rather than the full $L_1$ sensitivity bound, these results characterize the observed utility trend under the implemented Laplace scales. Based on this trend, $\epsilon=2$ and $\epsilon=10$ are selected as representative high-perturbation and lower-perturbation settings for comparison. 
In this table, $\epsilon$ indexes the implemented relation
$b_{\mathrm{impl}}=4.0/\epsilon$; the formal full-vector calibration is given separately in Equations~\ref{eq:laplace-formal-l1}--\ref{eq:laplace-formal-scales}. 

The results show that this perturbation setting preserves more statistical structure than the tested Gaussian configurations as $\epsilon$ increases. 
The results indicate that the implemented Laplace perturbation preserves more statistical utility than the analytically calibrated Gaussian mechanism in the evaluated configuration.
However, higher-utility settings also correspond to smaller perturbation magnitudes in the implemented sweep. 
Therefore, these results show a utility trend rather than establishing a generally optimal privacy--utility setting for clinical EEG anonymization.

\subsection{Machine Learning Model Assessment}
The downstream classification task was evaluated using Leave-One-Subject-Out (LOSO) validation at the patient level. In each fold, one patient was held out for testing, and the remaining patients were used for training. All reported classification metrics and confusion matrices are therefore computed from patient-level predictions, not from individual files, epochs, or windows. Window-level features were first aggregated into one feature representation per patient, and the class distribution used in the evaluation refers to these patient-level labels. 
Patient-level labels were assigned using the positive seizure-related keywords ``epileptic seizure,'' ``ictal EEG pattern,'' and ``ictal EEG activity.'' The expressions ``epileptiform interictal activity'' and ``abnormal interictal'' did not by themselves trigger a positive label. No additional automated negation handling or manual label verification was performed. 
This patient-level split prevents files, epochs, or windows from the same patient from appearing in both the training and test portions of a LOSO fold and remains consistent with the patient-level privacy unit used in the DP analysis. The downstream classifier was implemented using scikit-learn's \texttt{MLPClassifier} with hidden-layer sizes \texttt{(100)},
activation \texttt{relu}, solver \texttt{adam},
\texttt{alpha} = 0.0001, learning-rate setting \texttt{constant}, \texttt{max\_iter} = 200, \texttt{early\_stopping = False}, and
\texttt{random\_state = None}. Within each LOSO fold, \texttt{StandardScaler} was fitted on the training patients and applied to the held-out patient.

\begin{table}[!b]
\caption{Dataset configuration for the patient-level MLP evaluation.}
\centering
\begin{tabular}{|c|c|c|}
\hline
\textbf{Dataset} & \textbf{Evaluation} & \textbf{No. of Patients} / \textbf{Class Distribution (0/1)} \\ \hline
Original & LOSO & 17  /  12--5 \\ \hline
\end{tabular}
\label{tab:mlp_original_loso_result}
\end{table}

Table~\ref{tab:mlp_original_loso_result} shows the class distribution used in the patient-level evaluation, with 12 patients in class 0 and 5 patients in class 1. The epilepsy class was treated as the positive class ($y=1$), and the non-epilepsy class as the negative class ($y=0$).

For each feature representation, LOSO validation produced one predicted probability per patient. The decision threshold was selected from the aggregated LOSO predictions to maximize the F1-score for the epilepsy class. Because threshold selection and evaluation used the same cohort, the classification results are interpreted as cohort-level downstream utility comparisons rather than estimates of performance in an independent clinical population \cite{varma2006bias}. 
The same evaluation procedure was applied consistently to the original, Gaussian-perturbed, and Laplace-perturbed feature representations.  
For the classification comparison in Table~\ref{tab:combined_classification_results}, the Gaussian-perturbed features were generated using $\epsilon = 10.0$, $\delta = 10^{-5}$, $\Delta_2 = 4.0$, and $\sigma = 1.9996$, whereas the Laplace-perturbed features were generated using $\epsilon=10.0$, implemented sensitivity $4.0$, and $b=0.4$. The selected decision thresholds were $0.40$ for
the original features, $0.05$ for the Gaussian-perturbed features, and $0.10$ for the Laplace-perturbed features.

\subsubsection{Evaluation on Original EEG Data} 
Original EEG features were first used to establish a non-private baseline for epilepsy detection. Model performance was evaluated using accuracy, precision, recall, F1-score, and confusion matrix. 
As shown in Table~\ref{tab:combined_classification_results}, the MLP model produced a patient-level LOSO confusion matrix with TN = 9, FP = 3, FN = 2, and TP = 3 on the original EEG dataset. Based directly on this matrix, the model achieved an accuracy of 0.706, precision of 0.500, recall of 0.600, specificity of 0.750, balanced accuracy of 0.675, and F1-score of 0.545 for the epilepsy class. 
For comparison, a classifier that predicts every patient as non-epilepsy
would obtain the same overall accuracy of $12/17=0.706$, but a balanced
accuracy of 0.500. The original-feature MLP achieved a higher balanced
accuracy of 0.675 because it correctly classified patients from both
classes ($TN=9$ and $TP=3$), rather than predicting only the majority
class. 
These results provide a non-private baseline for the downstream utility check, but they should be interpreted cautiously due to the small and imbalanced dataset. In particular, cross-validation estimates obtained from such a small cohort can have large sampling variability and error bars \cite{varoquaux2018cv}.

\begin{table}[!b]
\centering
\caption{Patient-level LOSO classification performance across original, Gaussian-perturbed, and Laplace-perturbed EEG features.}
\begin{tabular}{|p{3cm}|c|c|c|c|c|c|c|}
\hline
\textbf{Dataset} & \textbf{Model} & \textbf{Accuracy} & \textbf{Precision} & \textbf{Recall} & \textbf{Specificity} & \textbf{Bal. Acc.} & \textbf{F1} \\ \hline
Original EEG & MLP & 0.706 & 0.500 & 0.600 & 0.750 & 0.675 & 0.545 \\ \hline
Gaussian-perturbed EEG & MLP & 0.294 & 0.294 & 1.000 & 0.000 & 0.500 & 0.455 \\ \hline
Laplace-perturbed EEG & MLP & 0.235 & 0.214 & 0.600 & 0.083 & 0.342 & 0.316 \\ \hline
\end{tabular}
\label{tab:combined_classification_results}
\end{table}

To quantify the uncertainty associated with the small patient cohort, two-sided 95\% Wilson score intervals were calculated directly from the patient-level confusion matrices. For the original features, the intervals are 0.469--0.867 for accuracy, 0.231--0.882 for recall, and 0.468--0.911 for specificity. The corresponding intervals are 0.133--0.531, 0.566--1.000, and 0.000--0.242 for the Gaussian-perturbed features, and 0.096--0.473, 0.231--0.882, and 0.015--0.354 for the Laplace-perturbed features. These intervals quantify the uncertainty associated with the limited cohort size. Because the decision thresholds were selected using the same cohort, the intervals are interpreted descriptively.

\begin{figure}[!t]
  \centering
  \subfloat[\centering Confusion matrix\label{fig:confusion_matrix_original}]{\includegraphics[width=0.5\linewidth]{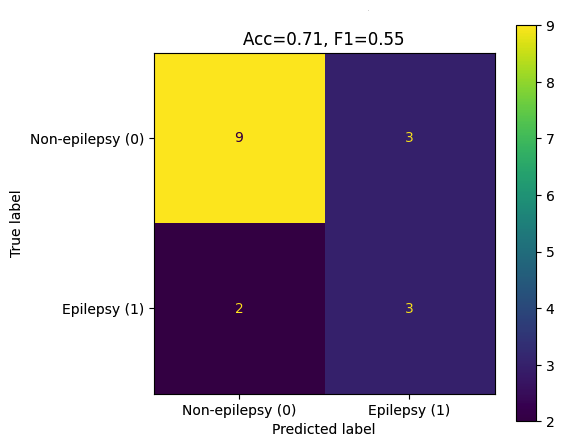}}
  \hfill
  \subfloat[\centering Accuracy and F1-score\label{fig:Loso_rgb_results}]{\includegraphics[width=0.5\linewidth]{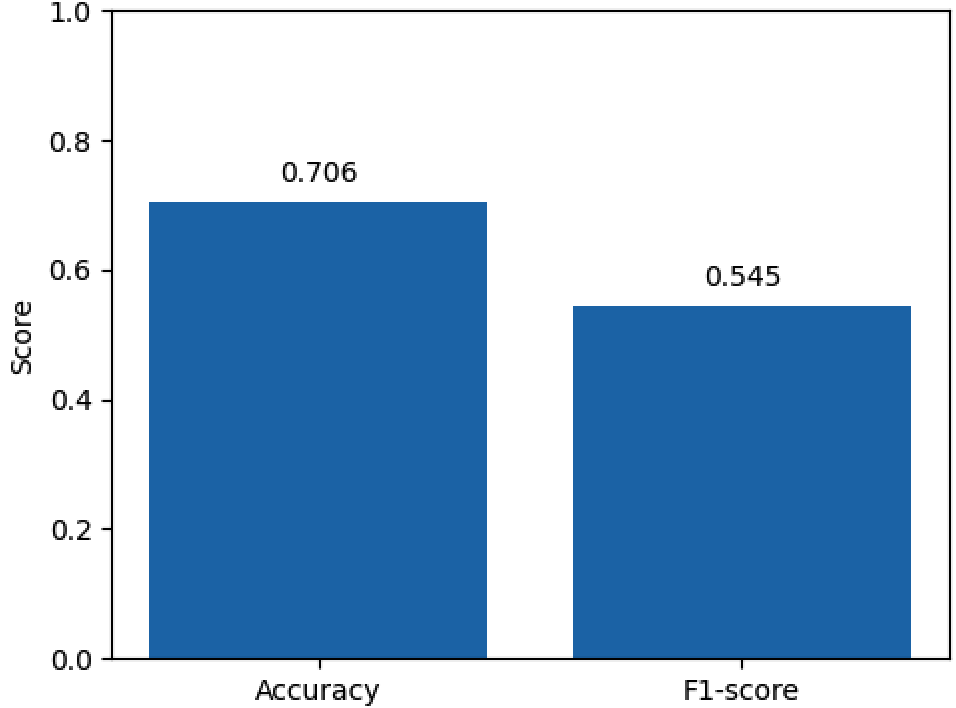}}
  \caption{MLP performance on original EEG features.}
\label{fig:original_mlp_results_combined}
\end{figure}

Figure~\ref{fig:confusion_matrix_original} shows the patient-level LOSO confusion matrix for the original EEG model. The model produced TN = 9, FP = 3, FN = 2, and TP = 3. These counts are consistent with the patient-level class distribution, since $TN+FP=12$ for the non-epilepsy patients and $FN+TP=5$ for the epilepsy patients.

\subsubsection{Evaluation on Gaussian-Perturbed EEG Data}

The Gaussian-perturbed features were evaluated using the patient-level LOSO procedure and threshold-selection approach described above. Accuracy, balanced accuracy, sensitivity, specificity, precision, F1-score, and the patient-level confusion matrix are reported. 
Table~\ref{tab:combined_classification_results} shows that the Gaussian-perturbed model obtained a low patient-level classification performance. Although recall was 1.000, specificity was 0, and precision was 0.294, meaning that all non-epilepsy patients were misclassified as epilepsy. Accordingly, the Gaussian-perturbed representation provides limited downstream classification utility because the model does not preserve class separability under this configuration.

\begin{figure}[!t]
  \centering
  \subfloat[\centering Confusion matrix\label{fig:gaussian_dp_mlp_cm}]{\includegraphics[width=.5\linewidth]{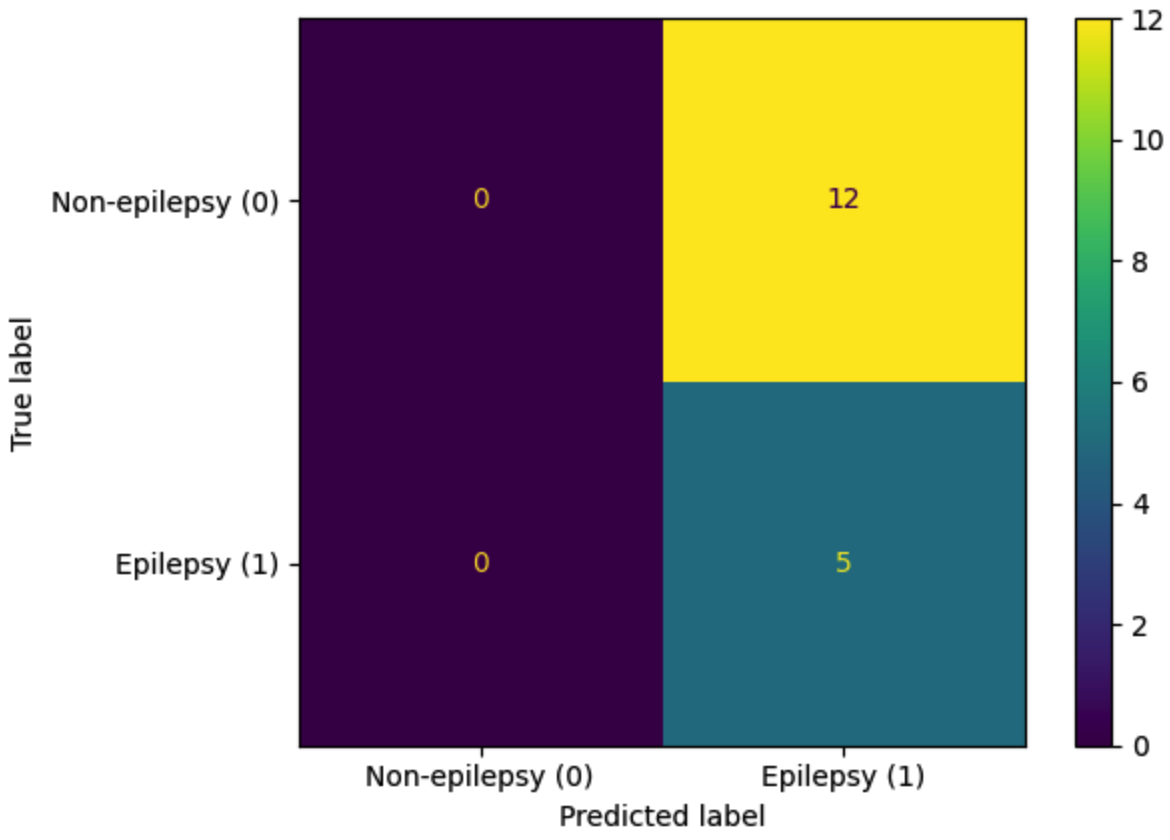}}
  \hfill
  \subfloat[\centering Accuracy and F1-score\label{fig:gaussian_dp_mlp_perf}]{\includegraphics[width=.5\linewidth]{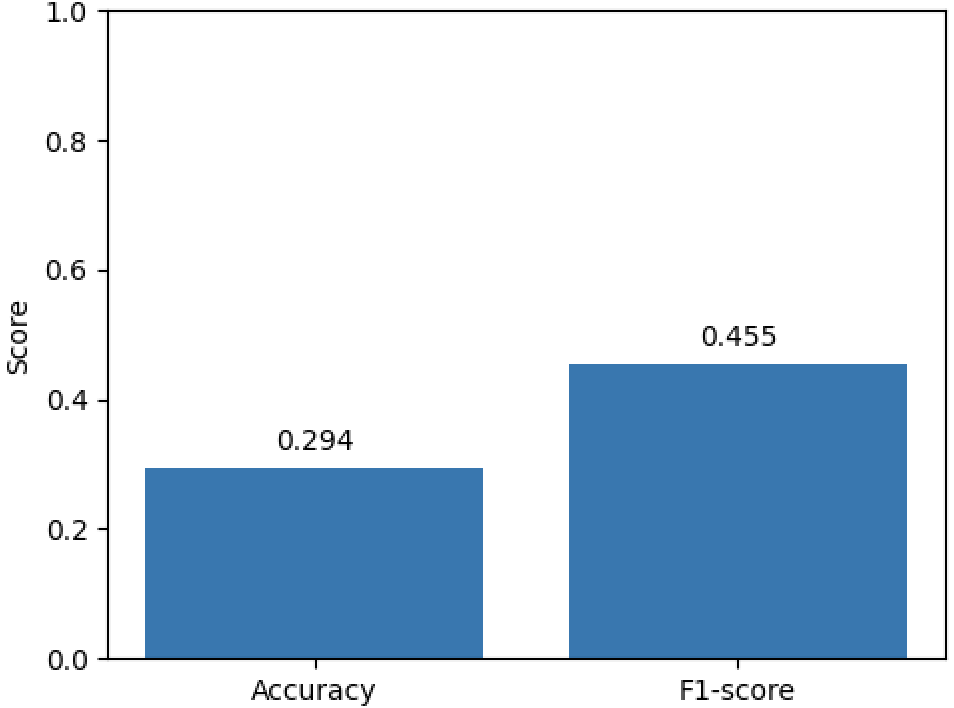}}
  \caption{MLP performance on Gaussian-perturbed EEG features.}
  \label{fig:gaussian_dp_results_combined}
\end{figure}

Figure~\ref{fig:gaussian_dp_mlp_cm} shows the confusion matrix for the MLP model trained on Gaussian-perturbed features. The model produced TN = 0, FP = 12, FN = 0, and TP = 5. Thus, all epilepsy cases were detected, but all non-epilepsy cases were incorrectly classified as epilepsy. Figure~\ref{fig:gaussian_dp_mlp_perf} summarizes the MLP performance on Gaussian-perturbed features. Because all patients were assigned to the positive class, the recall of 1.000 does not indicate clinically useful sensitivity. Instead, the zero specificity and all-positive prediction pattern demonstrate substantial loss of class separability under this Gaussian configuration.

\subsubsection{Evaluation on Laplace-Perturbed EEG Data}
The same LOSO evaluation was applied to the Laplace-perturbed patient-level EEG features generated in the experiments.
Probability-based predictions were used, and the decision threshold was tuned to maximize the F1-score for the epilepsy class. Table~\ref{tab:combined_classification_results} shows that the MLP model trained on Laplace-perturbed EEG features achieved limited patient-level classification performance by giving an accuracy of 0.235, precision of 0.214, recall of 0.600, specificity of 0.083, balanced accuracy of 0.342, and F1-score of 0.316. Although the model correctly detected three of the five epilepsy patients, it also misclassified eleven of the twelve non-epilepsy patients as epilepsy. The very low specificity and high false-positive count demonstrate substantial downstream utility degradation; this result should not be interpreted as evidence of clinical usefulness.

\begin{figure}[!t]
  \centering
  \subfloat[\centering Confusion matrix\label{fig:laplacedp_mlp_cm}]{\includegraphics[width=0.5\linewidth]{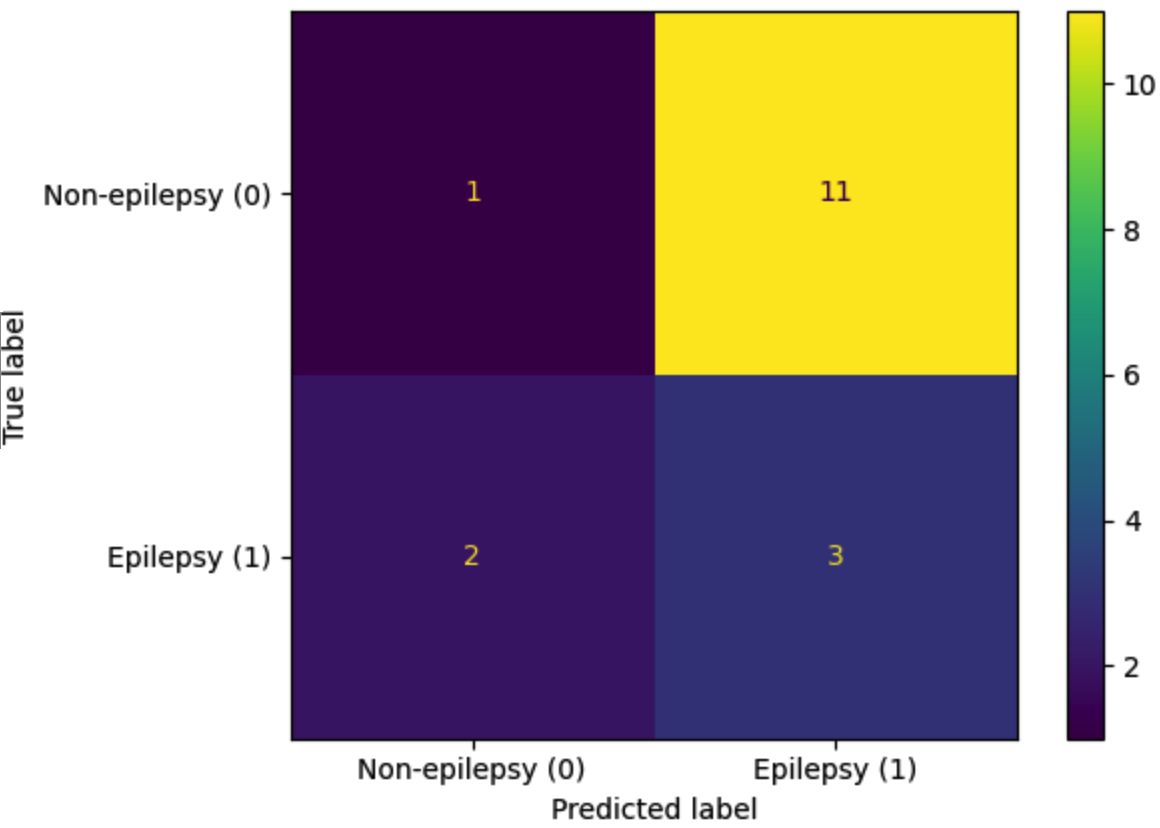}}
  \hfill
  \subfloat[\centering Accuracy and F1-score\label{fig:laplace_dp_mlp_perf}]{\includegraphics[width=0.5\linewidth]{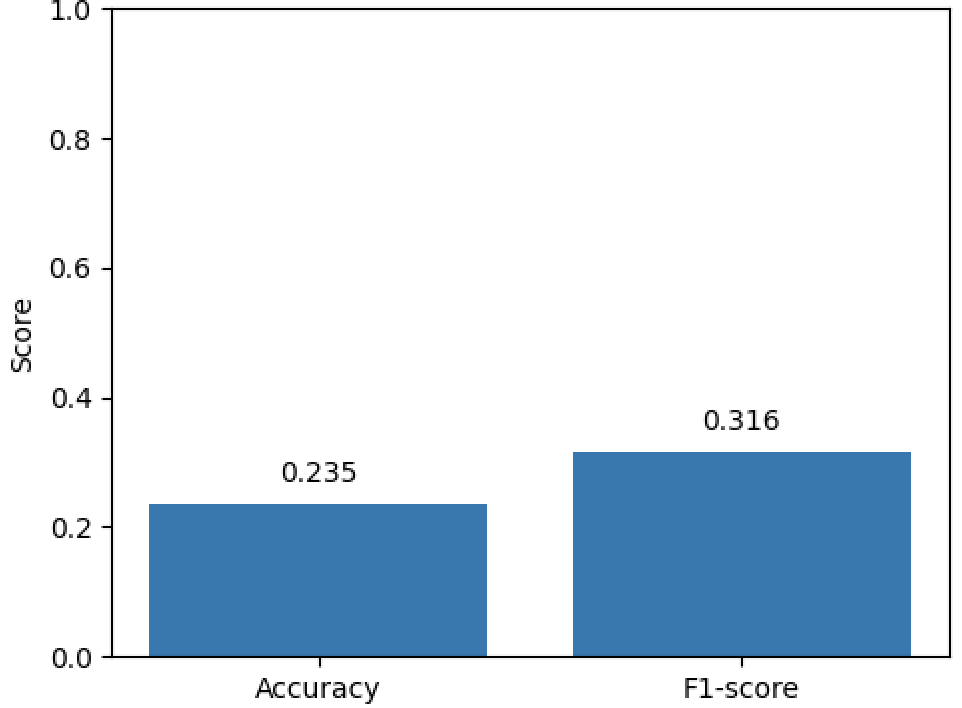}}
  \caption{MLP performance on Laplace-perturbed EEG features.}
  \label{fig:laplace_dp_results_combined}
\end{figure}

Figure~\ref{fig:laplacedp_mlp_cm} presents the confusion matrix for the MLP model trained on Laplace-perturbed features. The model produced TN = 1, FP = 11, FN = 2, and TP = 3. Compared with the Gaussian-perturbed setting, the implemented Laplace
perturbation produced one correctly classified non-epilepsy case but missed two epilepsy cases. Figure~\ref{fig:laplace_dp_mlp_perf} summarizes the corresponding accuracy and F1-score. Although the F1-score is slightly higher than the accuracy, the results remain weaker than the original-data baseline. This indicates that Laplace noise reduces feature separability and limits downstream classification performance.

Under the implemented settings, the Laplace perturbation preserves only limited classification utility for epilepsy detection and still produces a high number of false positives.
Compared with the Gaussian-perturbed setting, it slightly improves specificity
but also introduces false negatives. These results indicate that the Laplace-perturbed features have limited usefulness for classification in the
evaluated configuration.

\section{Conclusion}
Differentially private anonymization of clinical EEG-derived feature representations was studied in a multi-hospital setting. Three deployment scenarios were considered: client-side anonymization, centralized server-side anonymization, and decentralized local training. 
Following EEG preprocessing and feature extraction, Gaussian and Laplace perturbations were applied to the resulting patient-level EEG feature representations, and their impact was assessed using statistical utility measures and a downstream patient-level machine-learning utility check.
The Laplace experiments used the implemented scale
$b_{\mathrm{impl}}=4.0/\varepsilon$, while the full-vector $L_1$
calibration required for a formal Laplace guarantee was derived separately.

The results show that DP-based perturbation can be integrated into EEG processing workflows, while the selected mechanism, sensitivity calibration, privacy parameters, and evaluation unit strongly affect data utility. The Gaussian mechanism was analytically calibrated using the implemented $L_2$ sensitivity, with the normalization parameters treated as fixed during the perturbation stage. The Laplace experiments quantify utility under the implemented perturbation magnitudes, while the corresponding full-vector $L_1$ calibration is derived separately. In the evaluated small and imbalanced dataset, larger perturbation magnitudes introduced substantial distortion, and downstream classification performance was limited after anonymization. These findings support the practical integration of privacy-preserving perturbation mechanisms into clinical EEG processing, provided that the privacy unit, adjacency relation, sensitivity, noise calibration, and validation protocol are specified carefully. Future work should evaluate larger and more balanced EEG datasets, repeat experiments over multiple noise realizations, investigate end-to-end privacy accounting for additional data-dependent preprocessing choices, and include empirical privacy-risk analyses such as re-identification, linkage, membership inference, and reconstruction attacks.

\subsection*{Acknowledgment} 
This work was supported in part by the Dam Foundation under project number SDAM\_UTV532216 and by the Research Council of Norway under project number 331903. 

\subsection*{Ethics Statement} 
The collection and processing of data were approved by the Norwegian South-Eastern Regional Ethics Committee (REC; reference no.~689562) and the local data protection officer at Oslo University Hospital in accordance with applicable European Union and Norwegian requirements. The project was granted an exemption by REC from the requirement of informed consent. Only anonymized data were made available to the authors.

\bibliographystyle{IEEEtran}
\bibliography{Ref}
\end{document}